\documentstyle [12pt] {article}
\title  {TIME REVERSAL OF THE INCREASING GEOMETRICAL PROGRESSION OF THE POPULATION OF A SIMPLE BIOLOGICAL SPECIES }

\author{Vladan Pankovi\'c$^\ast$ , Rade Glavatovi\'c$^\sharp$ ,Milan Predojevi\'c $^\ast$ \\
$^\ast$Department of Physics , Faculty of Science, \\ 21000 Novi
Sad , Trg Dositeja Obradovi\'ca 4. ,
Serbia and Montenegro , \\
$^\sharp$Military Medical Academy \\11000 Beograd , Crnotravska 17
, Serbia and Montenegro \\
vladanp@gimnazija-indjija.edu.yu}
\date {}
\begin {document}
\maketitle

\vspace {1.5cm}

 \begin {abstract}
In this work we consider time reversal of the increasing
geometrical progression of the population of a simple biological
species without any enemies (predators) in the appropriate
environment with unlimited resources (food, territory, etc.). It
is shown that such time reversal corresponds to appearance of the
cannibalism, i.e. self-predaciousness or self-damping phenomena
which can be described by a type of difference Verhulst equation.
\end {abstract}
\vspace {1.5cm}

In this work we shall consider time reversal of the increasing
geometrical progression of the population of a simple biological
species without any enemies (predators) in the appropriate
environment with unlimited resources (food, territory, etc.). It
will be  shown that such time reversal corresponds to appearance
of the cannibalism, i.e. self-predaciousness or self-damping
phenomena which can be described by a type of difference Verhulst
equation. Also, it will be shown how such time reversal of the
population dynamics can be demonstrated formally in an environment
with changeable food concentration.

Suppose that an individual of a simple, monosexual biological
species (eg. a bacterium species), in an appropriate environment
with unlimited resources (food, territory, etc.) and without any
enemies (predators), during a reproduction time interval (cycle),
$\Delta t$, does a splitting (fision) in $k+1$ new individuals,
where k represents a natural number 1,2,3... Presented dynamical
change of the population (number of the individuals of given
species) by interaction with environment during one reproduction
time interval can be described by the following difference
equation
\begin {equation}
\frac {\Delta p}{\Delta t} = {\it a}p_{in}
\end {equation}
Here a is positive and it represents a constant that describes
phenomenologically positive interaction between individual and
environment. (By positive interaction between individual and
environment there is increase of the population.) It can be
presented in the following form
\begin {equation}
{\it a} = \frac {k}{\Delta t}
\end {equation}
Also,
\begin {equation}
\Delta p = p_{fin} - p_{in}
\end {equation}
where $p_{in}$ and $p_{fin}$  represent initial (at the beginning
of the reproduction cycle) and final (at the end of the
reproduction cycle) population in the general solution of (1). As
it is not hard to see given general solution of (1), i.e. of
\begin {equation}
\frac {\Delta p}{\Delta t} = \frac {k}{\Delta t}p_{in}
\end {equation}
is
\begin {equation}
p_{fin} = (k+1) p_{in}
\end {equation}
which represents a geometrical progression with coefficient $1+k$
greater than 1. Supposed especial solution of (1), i.e. (4) that
satisfies initial condition
\begin {equation}
p_{in} = 1
\end {equation}
is
\begin {equation}
p_{fin} = k+1
\end {equation}

Population dynamical equation (1), i.e. (4)  holds a very
important characteristics that can be called {\it locality}.
Namely, population change during reproduction time interval (left
hand-side of (1), i.e. (4) ) represents a function (right
hand-side of (1), i.e. (4)) that depends of the initial population
and that does not depend of the final population. Or, given
population dynamics does not need that final population is known a
priori, but this final population can be determined by initial
population and population dynamical equation (1), i.e. (4). Simply
speaking, there is a finite speed of the population dynamical
evolution form the past toward future. In case that population
change during reproduction time interval in a population dynamical
equation  represents a function  that depends of the final
population too it can be said that corresponding population
dynamics is non-local or instantaneous. Such population dynamics
needs, in fact, that both initial and final population are known
simultaneously. But such non-local population dynamics does not
exist really within ecology.

It is not hard to see that locality of the population dynamics
corresponds conceptually to locality of the physical dynamics.
Simply speaking locality of the physical dynamics means in fact
that there is no physical interaction faster than speed of light.

Now, apply time reversal transformation, $T$, [1], [2] that
changes discretely time moment $t$ in $t' = - t$ , or initial in
the final conditions in the general solution of an equation, at
(1), i.e. (4). It yields
\begin {equation}
\frac {p'_{in} - p'_{fin}}{\Delta t} = \frac {k}{\Delta t}p'_{fin}
\end {equation}
where p'in and p'fin  represent initial (at the beginning of the
reproduction cycle) and final (at the end of the reproduction
cycle) population in the general solution of (8). As it is not
hard to see given general solution is
\begin {equation}
p'_{fin} = \frac {1}{k+1}p'_{in}
\end {equation}
which represents a geometrical progression with coefficient
1/(1+k) smaller than 1. For especially chosen
\begin {equation}
p'_{in} = k+1
\end {equation}
it follows, as an especial solution of (8),
\begin {equation}
p'_{fin} = 1
\end {equation}

It is not hard to see that geometrical progression (9) represents
generally the inversion of the geometrical progression (5), so
that, in the especial case, it follows that (6) is numerically
equivalent to (11)  and (7) to (10). But, equation (8) can be
transformed in the equivalent equation
\begin {equation}
\frac {\Delta p'}{\Delta t} = -\frac {k}{\Delta t}p'_{fin} = -{\it
a}p'_{fin}
\end {equation}
Here {\it -a} is negative and it represents a constant that
describes phenomenologically negative interaction between
individual and environment. (By negative interaction between
individual and environment there is decrease of the population.)
Also, here
\begin {equation}
\Delta p' = p'_{fin} - p'_{in}
\end {equation}
But, in distinction from (1), i.e. (4), equation (12) is
non-local. For this reason equation (12) representing time
reversal of (1), i.e. (4)  has not real ecological sense.

Suppose however that splitting of the individual during one
reproduction cycle is recorded by a video camera and that later a
play back backward is done. It formally corresponds to application
of $T$ at (1), i.e. (4). Then one can observe how at the
"beginning" there are $k+1$ individuals which interact mutually
during time interval $\Delta t$ so that only one individual stands
while $k$ other individuals disappear at the "end" of this
interval. Simply speaking, effectively, one of $k+1$ initial
individuals kills (eats) all $k$ other individuals. In this way
here an effective cannibalism, i.e. self-predaciousness or
self-damping phenomena within given species appears. If given
species is sufficiently simple then cannibalism within given
species can be biologically  admirable.

Corresponding population dynamics that includes cannibalism , i.e.
self-predaciousness or self-damping appearance, can be presented
by the following local difference equation
\begin {equation}
\frac {\delta p'}{\Delta t} = \frac {k}{\Delta t}p'_{in}  - \frac
{{\it b}}{\Delta t}(p'_{in})^{2}
\end {equation}
where ${\it b}$ represents self-damping constant. Obviously, (14)
represents a type of the difference Verhulst equation [3]-[8].
Self-damping constant ${\it b}$ can be determined from (14), (13)
and conditions (10), (11). It yields
\begin {equation}
b = \frac {k(k+2)}{(k+1)^{2}}
\end {equation}

Thus, both equations (1), i.e. (4) and (14) have, at least
effectively, real ecological sense. Also (14) can be consistently
effectively treated as time reversal of (1), i.e. (4), and vice
versa. For this reason it can be consistently stated that set of
given two equations (1), i.e. (4) and (14) is $T$ invariant
(symmetric) or time reversible (even if neither (1), i.e. (4) nor
(14) is time reversible).

But, as it is well-known, time reversal does not correspond to any
real ecological interaction. Nevertheless, instead of time
reversal an analogous discrete transformation of (1), i.e. (4)
with equivalent consequences can be done.

Namely, suppose that b represents a discrete function of the food
concentration, $C$, etc. in the environment so that
\begin {equation}
b = 0      \hspace{1cm}  {\rm for} \hspace{1cm}C \geq C_{0}
\end {equation}
\begin {equation}
b = \frac {k(k+2)}{(k+1)^{2}}  \hspace{1cm} {\rm for} \hspace{1cm}
C < C_{0}
\end {equation}
where $C_{0}$ represents the critical value of $C$. In this case
solution of (14) under condition (16) and initial condition
\begin {equation}
p'_{in} = 1
\end {equation}
equivalent to (6) is
\begin {equation}
p'_{fin} = k+1
\end {equation}
equivalent to (7), while, numerically inversely,  solution of (12)
under condition (17) and initial condition (10) is equivalent to
(11).

It can be very interesting that theoretical results (14)-(17) be
compared with empirical data, i.e. with results of the
experimental analysis of the critical value of the food
concentration in environment for simple biological species (eg.
bacterium species etc.) that admits more or less discrete
appearance of the cannibalism for food concentration smaller than
critical. Especially it can be interesting for
\begin {equation}
k = 1
\end {equation}
when, according to (19),
\begin {equation}
b = 3/4  = 0.75
\end {equation}
In conclusion the following can be repeated and pointed out.
Population dynamics of a simple biological species, living without
any enemies, in an appropriate environment with unlimited
resources, corresponding  to increasing geometrical progression,
turns by time reversal in a population dynamics corresponding to
difference Verhulst population dynamics. All this can be
demonstrated formally in an environment with changeable food
concentration.

\section {References}

\begin {itemize}

\item [[1]]  N.N.Bogolubov, A.A.Logunov, T.Todorov, {\it Introduction to Axiomatic Quantum Field Theory} (W.A.Benjamin, Reading, Mass., 1975.)
\item [[2]]  R.F.Streater, A.S.Wightman, {\it PCT , Spin and Statistics and All That} (Addison-Wesley, New York, 1989.)
\item [[3]]  R.M.May, {\it Stability and Competition in Model Ecosystems} (Princeton Univ.Press., Princeton, New Jersey,1974.)
\item [[4]]  E.C.Pielou, {\ it An Introduction to Mathematical Ecology} (John Wiley and Sons, New York, 1969.)
\item [[5]]  E.C.Pielou, {\it Mathematical Ecology} (John Wiley and Sons, New York, 1977.)
\item [[6]]  F.Verhulst, {\it Nonlinear Differential Equations and Dynamical Systems} (Springer Verlag, Berlin, 1990.)
\item [[7]]  {\it Foundations of Ecology: Classical Papers with Commntaries}, eds.  L.A.Real, J.H.Brown (University of Chicago Press, Chicago, Illinois, 1991.)
\item [[8]]  J.D.Murray, {\it Mathematical Biology} (Springer Verlag, Berlin-Heidelberg, 1993.)

\end {itemize}

\end {document}